\documentstyle[prb,aps,multicol,epsf]{revtex}
\twocolumn
\begin{document}
\draft
\title{Universal quantum gates based on both geometric and dynamic 
phases in quantum dots}
\author{Kaiyu Yang$^{1}$, Shi-Liang Zhu$^{1,2}$
\thanks{Email address: shilzhu@yahoo.com.cn;
Tel.: (852) 2859 2361; Fax: (852) 2559 9152},
and Z. D. Wang$^{1,3}$
}
\address{$^1$Department of Physics, University of Hong Kong, Pokfulam Road, Hong Kong, China. \\
$^2$ Department of Physics,
South China Normal University, Guangzhou, China\\
$^3$ Department of Material Science and Engineering,
University of Science and Technology of China, Hefei, China
}
\address{\mbox{}}
\address{\parbox{14cm}{\rm \mbox{}\mbox{}
A large-scalable quantum computer model, whose qubits are represented by the
subspace subtended by the ground state and the single exciton state on
semiconductor quantum dots, is proposed. A universal set of quantum gates in
this system may be achieved by a mixed approach, composed of dynamic
evolution and nonadibatic geometric phase.
}}
\address{\mbox{}}
\address{\parbox{14cm}{\rm PACS:
03.67.Lx, 03.65.Vf,85.35.Be, 71.35.Cc
}}

\maketitle
\newpage
\narrowtext

Quantum computation has been demonstrated to be more powerful than classical
computation by the algorithm of the prime factoring of an integer.\cite{Shor}
Recently great interest has been sparked into the research of quantum
computation. One of the focus points is how to construct an experimentally
feasible quantum computer to do quantum computation task. A universal set of
quantum gates, \cite{D. Deutsch,S. Lloyd} composed of two non-commutable
single-qubit (quantum bit) gates and one non-trivial two-qubit gate, are
sufficient to implement an arbitrary quantum operation on $n$ qubits. \cite%
{Nielsen} How to construct a universal set of quantum gates based on
experimentally suitable physical systems has become a hot topic in the field
of quantum computation.

So far, a quite number of quantum computer protocols have been proposed
based on trapped ions, cavity quantum electrodynamics, superconductor
Josephson junctions, and nuclear magnetic resonance (NMR) schemes. \cite%
{Nielsen} With the conditions for quantum computation so extremely
demanding, decoherence, operation error and lack of capability to be
large-scalable have been the main obstacles for their experimental
realization.\cite{Nielsen}

With the recent development of nanosemiconductor fabrication and
manipulation, solid-state based semiconductor quantum dots (QDs) have shown
to be a promising candidate for quantum computation. The nature of being
solid-state based makes QDs large-scalable. A number of proposals have been
presented based on QDs either by using the spins of net charges in QDs, the
nuclear spins of donor atoms to construct quantum computers, \cite{A.
Imamoglu 2,Paolo Solinas} or by exploiting exciton state space for quantum
information processing.\cite{Luis Quiroga} Nevertheless, little work has
been done in the implementation of a universal set of quantum gates in the
subspace subtended by the ground and single exciton states, at least
explicitly. In this paper, we show that the above mentioned system may be a
practical candidate for large-scalable and fault-tolerant quantum computers.
The atomiclike discrete excitonic energy distribution makes QDs highly
excluded from environment decoherence. Ultrafast manipulation and coherent
control of excitons on one single quantum dot and high spatially resolved
microprobing have been accomplished experimentally.\cite{N.H. Bonadeo 2,H.
Kamada} The experimental observation of Rabi oscillations from excitons
confined to single QDs was reported,\cite{Stievater} which corresponds to
realize a single-qubit rotation. Furthermore, optically induced entanglement
is experimentally identified by the spectrum of the phase sensitive
homodyne-detected coherent nonlinear optical response.\cite{Gang Chen
Science} These experiments demonstrate that it is possible to manipulate
quantum dot excitons by current technology, which is important for quantum
computation. Motivated by these developments, we propose an experimentally
feasible scheme to accomplish a set of universal quantum gates in quantum
dot excitons.


To achieve a universal set of quantum gates, two quantum mechanical methods
are available, one by dynamic evolution and the other by geometric operation
adiabatically or nonadiabatrically. Dynamic evolution is utilized more often
due to its more direct and simpler operation process and higher adaptability
to different real systems. But this method is vulnerable to decoherence from
environment noise. Geometrical phase has been proposed to be a
fault-tolerant approach because it depends only on the solid angle enclosed
by the parameter path in adiabatic cases or by the evolution curve in the
projective Hilbert space in cyclic cases. Geometrical quantum computation %
\cite{Zanardi} based on the adiabatic geometric phase has been presented
using superconducting nanocircuits, \cite{G. Falci} NMR, \cite{Jonathan A.
Jones} and trapped ions.\cite{L.M. Duan} However, the adiabatic condition in
these schemes requires long coherent time to complete the evolution, which
is a great challenge for current solid-state technology. A specific
nonadiabatic operation was proposed to evade such tight restriction.\cite%
{Wang Xiangbin 4} Furthermore, a general scheme for implementation of
universal quantum gates based on nonadiabatic geometric phases has also been
proposed.\cite{Zhu} In this paper, a mixed scheme, composed of nonadiabatic
geometric phase and dynamic evolution, is worked out to achieve a universal
set of quantum gates, which to some extent can be fault-tolerant and
large-scalable.

We consider a system of $N$ identical and equispaced quantum dots similar to
that in Ref. \cite{Luis Quiroga}, which is radiated by long-wavelength
classical light without net charges contained in each dot. As we restrict at
most only one exciton is possible to exist in each quantum dot, the
formation of single exciton within the corresponding individual QDs and the
interdot Coulomb exchange (F\"{o}rster) interaction between a pair of
coupled dots can be described in the framework of the rotating-wave
approximation (RWA) by the Hamiltonian, ($\hbar =1$),\cite{Luis Quiroga}
\begin{eqnarray}
H(t) &=&\frac{\epsilon }{2}\sum_{i=1}^{N}\{c_{i}^{+}c_{i}-h_{i}h_{i}^{+}\}
\nonumber \\
&&-\frac{V}{2}\sum_{{%
{i,j=1 \atop i\neq j}%
}}^{N}(c_{i}^{+}h_{j}c_{j}h_{i}^{+}+h_{i}c_{j}^{+}h_{j}^{+}c_{i})  \nonumber
\\
&&+E(t)\sum_{i=1}^{N}c_{i}^{+}h_{i}^{+}+E^{\ast }(t)\sum_{i=1}^{N}h_{i}c_{i},
\label{Hamiltonian 1}
\end{eqnarray}%
where $c_{i}^{+}(c_{i})$ is the electron creation (annihilation) operation
in the conductive band of the $i$th quantum dot, $h_{j}^{+}(h_{j})$ is the
creation (annihilation) operation for hole in the valent band of the $j$th
quantum dot, $\epsilon $ is the energy band gap of semiconductor dots with
the middle of the gap being set as zero-energy point, $V$ denotes the
interdot interaction inducing the transference of an exciton from one QD to
the other
and can be experimentally controlled very well in a relative
large regime from zero coupling to the strong coupling regime
by using a split-gate technique, \cite{Livermore}
and $E(t)$\ describes the effective laser pulse shape.
Here we have assumed that $\epsilon $ in different dots are the same.
Although $\epsilon $ for different QDs fabricated by self-assembled method
are not exactly equal, this assumption is still reasonable as
its value may be
slightly modified by a bias voltage $V_{b}$ independently acting on each dot.
\cite{Zrenner}
The operators involved in Eq.(\ref{Hamiltonian 1}) obey the
anticommutation rules
$\{c_{i}^{+},c_{j}\}=\{h_{i}^{+},h_{j}\}=\delta _{i,j}$.

By introducing a new set of operators $\{J_{i+}=c_{i}^{+}h_{i}^{+}$, $%
J_{i-}=h_{i}c_{i},$ $J_{iZ}=(c_{i}^{+}c_{i}-h_{i}h_{i}^{+})/2 \}$ satisfying
quasi-spin relations: $[J_{i+},J_{j-}]=2\delta _{ij}J_{iZ},[J_{i\pm }$, $%
J_{jZ}]=\mp \delta_{ij}J_{\pm i}$, Eq. (\ref{Hamiltonian 1}) can be
rewritten as\cite{Luis Quiroga}
\begin{eqnarray}
H(t) &=&\epsilon \sum_{i=1}^{N}J_{iZ}-\frac{V}{2} \sum_{{%
{ i,j=1  \atop i\neq j}%
}}^{N}(J_{i+}J_{j-}+J_{i-}J_{j+})  \nonumber \\
&&+E(t)\sum_{i=1}^{N}J_{i+}+E^{\ast }(t)\sum_{i=1}^{N}J_{i-}.
\label{Hamiltonian 2}
\end{eqnarray}
>From this Hamiltonian we start to implement a universal set of quantum gates.

We first address how to realize the single-qubit gate. For single quantum
dot, the Hamiltonian in Eq.(\ref{Hamiltonian 2}) becomes:
\begin{equation}
H(t)=\epsilon J_{Z}+E(t)J_{+}+E^{\ast }(t)J_{-}.  \label{Hamiltonian 3}
\end{equation}
We choose the ground state without exciton and the single exciton state as
the basis $\{ | 0 \rangle, | 1\rangle \}$ of the computation Hilbert space
of the qubit. By introducing another set of quasi-Pauli operators $\{
X=J_{+}+J_{-}$, $Y=i(-J_{+}+J_{-})$, $Z=2J_{Z} \}$ with the commutation
relations $\{ [Z,X]=2iY$, $[Y,Z]=2iX,$ $[X,Y]=2iZ\}$ and choosing the
incident laser pulse shape $E(t)=A \exp(i\omega t)$ with $\{A,\omega \}$ the
\{amplitude, frequency\} of the laser respectively, the Hamiltonian in Eq.(%
\ref{Hamiltonian 3}) takes the form%
\begin{equation}
H(t)=\frac{\epsilon }{2}Z+A\cos (\omega t)X+A\sin (\omega t)Y.
\label{single qubit}
\end{equation}

In order to make quantum gates fault-tolerant to some extent, we implement
nonadiabatic geometric operation to achieve single qubit gates in our
system. In the projective Hilbert space--a unit sphere S$^{2\text{ }}$%
subtended by $\{\left| 0\right\rangle $, $\left| 1\right\rangle \}$, by
taking the incident pulse of laser as the rotation field, and the
semiconductor energy gap as the constant z-component field, there always
exist a pair of orthogonal states $|\psi _{\pm }\rangle $ which can evolve
cyclically and be expressed as $\{|\psi _{+}\rangle =\cos (\chi /2)|0\rangle
+\sin (\chi /2)|1\rangle$, $|\psi _{-}\rangle =-\sin (\chi /2)|0\rangle
+\cos (\chi /2)|1\rangle \}$ with $\chi =atan[2A/(\epsilon -\omega )]$. By
taking into account of the cyclic evolution, we have one simple relation $%
U(\tau )|\psi _{\pm }\rangle =\exp (\pm i\gamma )|\psi _{\pm }\rangle $,
where $\pm \gamma $ are the total phases accumulated in the cyclic evolution
of the orthogonal states $|\psi _{\pm }\rangle $ respectively. For an
arbitrary initial state as $\left| \Psi _{i}\right\rangle =a_{+}|\psi
_{+}\rangle +a_{-}|\psi _{-}\rangle $, the final state after a cyclic
evolution of time $\tau =2\pi /\omega $ is found to be $\left| \Psi
_{f}\right\rangle =U(\chi ,\gamma )\left| \Psi _{i}\right\rangle $, where $%
U(\chi ,\gamma )$ is given by\cite{Zhu}
\[
\left(
\begin{array}{ll}
e^{i\gamma }\cos ^{2}\frac{\chi }{2}+e^{-i\gamma }\sin ^{2}\frac{\chi }{2} &
i\sin \chi \sin \gamma \\
i\sin \chi \sin \gamma & e^{i\gamma }\sin ^{2}\frac{\chi }{2}+e^{-i\gamma
}\cos ^{2}\frac{\chi }{2}%
\end{array}%
\right) . 
\]
Even though $U(\chi ,\gamma )$ is achieved by the phase composed of
nonadiabatic geometric phase and dynamic phase, we can remove the dynamic
phase using the methods described in Ref.\cite{Zhu_quant_ph0210027}, and the
geometric quantum gates may be realized. For different values of $\{\chi
,\gamma \}$, $U(\chi _{1},\gamma _{1})$ and $U(\chi _{2},\gamma _{2})$ are
noncommutable except for the case $\sin \gamma _{1}\sin \gamma _{2}\sin
(\chi _{2}-\chi _{1})=0$. Since $\{\chi ,\gamma \}$ depend only on the
amplitude of the incident pulse of laser which can be adjusted as a
continuum, and the symmetric axis of the Hamiltonian in Eq. (\ref{single
qubit}) can be modified by changing the effective field, we are able to
achieve arbitrary single qubit gate. As this kind adjustment is relative
easy and convenient, such scheme is experimentally feasible. Two well known
universal single-qubit gates defined by $U_{Z}(\gamma_{z})=U(0,-\gamma_z/2)$
and $U_{X}(\gamma _{x})=U(\pi/2,-\gamma/2)$ may be realized, where $-\gamma
_{x,z}/2$ are geometric phases accumulated in the evolution. \cite{Zhu}

We then address the implementation of two-qubit gates in the system. Two
coupled QDs without the incident laser are considered. The coupling between
the two coupled QDs plays the key role to generate entanglement. Using the
set of quasi-Pauli operators $\left\{ X,Y,Z\right\} $ introduced previously,
the Hamiltonian described in Eq.(\ref{Hamiltonian 2}) now takes the form\cite%
{Luis Quiroga}
\begin{equation}
H(t)=\frac{\epsilon }{2}(Z_{1}\otimes I_{2}+I_{1}\otimes
Z_{2})-V(X_{1}\otimes X_{2}+Y_{1}\otimes Y_{2}),  \label{no light}
\end{equation}%
with $I$ a $2\times 2$ unit matrix. For this specific physical system, it
does not seem to be feasible to work out a two-qubit gate by conditional
geometric phase shift, and thus we use dynamic evolution to achieve a
two-qubit $iSWAP$ \ gate. In the computation space subtended by $\{\left|
00\right\rangle ,\left| 01\right\rangle ,\left| 10\right\rangle ,\left|
11\right\rangle \}$, Eq.(\ref{no light}) can be rewritten in a matrix form
as
\begin{equation}
H=\left(
\begin{array}{cccc}
\epsilon  & 0 & 0 & 0 \\
0 & 0 & -2V & 0 \\
0 & -2V & 0 & 0 \\
0 & 0 & 0 & -\epsilon
\end{array}%
\right) .  \label{coupled}
\end{equation}%
With the eigenstates \{$|00\rangle ,(|01\rangle +|10\rangle )/\sqrt{2}%
,(|01\rangle -|10\rangle )/\sqrt{2},|11\rangle $\}, the evolution operation
is derived exactly as
\[
U^{(1,2)}(t)=\left(
\begin{array}{cccc}
e^{-it\epsilon } & 0 & 0 & 0 \\
0 & \cos (-2Vt) & i\sin (-2Vt) & 0 \\
0 & i\sin (-2Vt) & \cos (-2Vt) & 0 \\
0 & 0 & 0 & e^{it\epsilon }%
\end{array}%
\right) .
\]%
Simply by setting $t\epsilon =2k\pi $ and $Vt=m\pi -\pi /4$ with $k,m$
positive integers, we obtain an appropriate elementary two-qubit $iSWAP$
gate $U^{iS}$ as
\begin{equation}
U^{iS}=\left(
\begin{array}{cccc}
1 & 0 & 0 & 0 \\
0 & 0 & i & 0 \\
0 & i & 0 & 0 \\
0 & 0 & 0 & 1%
\end{array}%
\right) .  \label{iswap}
\end{equation}%
By applying the $iSWAP$ gate twice, the famous $CNOT$ operation $U^{CN}$ can
be achieved \cite{Norbert Schuch}
\begin{eqnarray}
U^{CN} &=&U_{Z}^{(2)}(-\frac{\pi }{2})U^{iS}U_{X}^{(1)}(\frac{\pi }{2})U^{iS}
\nonumber \\
&&U_{Z}^{(1)}(\frac{\pi }{2})U_{Z}^{(2)}(-\frac{\pi }{2})U_{X}^{(2)}(-\frac{%
\pi }{2}).  \label{CNOT}
\end{eqnarray}
(up to an irrelevant overall phase) where $j$ in $U^{(j)}$ ($j=1,2$) denotes
the single-qubit gate acting on the $j$th qubit.  Therefore we have
presented a scheme to achieve a universal set of quantum gates successfully.
It is worth pointing out that additional phase shifts
should be taken into account because the eigenstates of the Hamiltonians
presented here are nondegenerate even at idle periods. This is one
common issue implied in many quantum computer models and may be solved
by controlling time with high accuracy concerning the start
time and time span of any quantum operation.\cite{Makhlin}

Considerable research efforts have been paid to design quantum computers
using QDs. Some schemes \cite{A. Imamoglu 2,Paolo Solinas} based on spin
manipulation in QDs generally have the advantage of the long decoherence
time of spin freedom with a upper limit up to 1-100$\mu s$, which is much
longer than that of charge freedom,\cite{Paolo Solinas}
However, these schemes often depend on the deliberately chosen isotropic
spin-spin interaction models while an anisotropic interaction induced by
spin-orbit coupling is inevitably present in a practical system. Also a
scheme to make use of anisotropic interaction was proposed, but additional
pulses with a modest constant factor are necessary.\cite{Lian-ao} Some other
schemes \cite{C. Piermarocchi,S. De Rinaldis} have pursued to implement
quantum information processing in a subspace subtended by the ground, single
exciton and multiexciton states in single QD, and take this subspace as a
two-qubit information register. However, to combine two logic qubits with
absolutely different state spaces into one physical register might be
neither large-scalable nor convenient to transfer and manipulate quantum
information. Furthermore, as the multiexciton state is of higher energy,
high excitation density is needed. It may lead to a shorter dephasing and
relaxation time due to the phonon-scattering and delocalization of exciton
states. Being different from them, in our scheme, the subspace is only
subtended by the ground state and the single exciton state. Thus the states
in every quantum dot is homogeneous, and the interaction is an effective
isotropic one. The dephasing time $\tau _{d}$ of the single exciton state is
about 40ps.\cite{H. Kamada}
The number of operation that can be completed within $\tau _{d}$ may
be estimated as follows. The common parameter $V$ ( $\epsilon $ ) is about
$0.1$eV ( $1.4$ eV ) for GaAs QDs,\cite{Luis Quiroga,F.J. Rodriguez}
and the operation span is mainly determined from the longer time factor $\tau
_{v}$ which is about $\tau _{v}\sim \hbar /V\sim 10fs$. Thereby the number
of gate operations should be $\tau _{d}/\tau _{v}\sim 10^{3}$, which is
large enough to demonstrate some typical quantum algorithms. As the highly localized
laser pulse, the external field and the coupling between any two neighboring quantum
dots can be controlled independently,
the extension of our scheme to an array of N quantum
dots should be possible. Moreover, at the end of the quantum computation,
the readout of a single exciton in QD can be achieved with high detection
efficiency by photoluminescence and various high resolution microprobing
methods. All these make our scheme experimentally feasible.

In conclusion, we propose a feasible scheme to construct a promising quantum
computer. The advantages of our scheme are: decoherence effects may
partially be suppressed by geometric quantum gates, and the
large-scalability for quantum computation may be accomplished from QDs
inherent solid-state feature.

This work was supported by the RGC grant of Hong Kong under Grants Nos.
HKU7118/00P and HKU7114/02P, and
a CRCG grant of HKU. S. L. Z. was supported in part
by SRF for ROCS, SEM, the NSF of Guangdong
under Grant No. 021088, and the
NNSF of China under Grant No. 10204008.


\section*{References}
\begin{thebibliography}{99}

\bibitem{Shor} Shor P W 1994 {\it  Proceedings 35$^{th}$ annual Symposium on
Foundations of Computer Science} (IEEE Press, Los Alamitos, CA)

\bibitem{D. Deutsch} Deutsch D, Barenco A and Ekert A
1995 {\it Proc. R. Soc.} A \textbf{449} 669

\bibitem{S. Lloyd} Lloyd S
1995 {\it Phys. Rev. Lett.} \textbf{75} 3461.

\bibitem{Nielsen} Nielsen M A and Chuang I 2000 {\it Quantum Computation and
Quantum Information} (Cambridge University Press, Cambridge, England)

\bibitem{A. Imamoglu 2} Imamoglu A,
Awschalom D D, Burkard G, Divincenzo D P, Loss D, Sherwin M and Small A
1999 {\it Phys. Rev. Lett.} \textbf{83} 4204

\bibitem{Paolo Solinas} Solinas P,
Zanardi P, Zanghi N and Rossi F
2002 {\it Preprint} quant-ph/0207019

\bibitem{Luis Quiroga} Quiroga L and Johnson N F
1999 {\it Phys. Rev. Lett.} \textbf{83} 2270

\bibitem{N.H. Bonadeo 2} Bonadeo N H,
Erland J, Gammon D, Park D, Katzer D S and Steel D G
1998 {\it Science} \textbf{282} 1473

\bibitem{H. Kamada} Kamada H, Gotoh H, Temmyo J, Takagahara T and Ando H
2001 {\it Phys. Rev. Lett.} \textbf{87} 246401

\bibitem{Stievater} Stievater T H, Li X, Steel D G, Gammon D, Katzer D S, Park D, Piermarocchi C and Sham L J
2001 {\it Phys. Rev. Lett.} \textbf{87} 133603

\bibitem{Gang Chen Science} Chen G, Bonadeo N H, Steel D G, Gammon D, Katzer D S, Park D and Sham K J
2000 {\it Science} \textbf{289} 1906

\bibitem{Zanardi} Zanardi P and Rasetti M
1999 {\it Phys. Lett.} A \textbf{264} 94

\bibitem{G. Falci} Falci G, Fazio R, Palma G M, Siewert J and Vedral V
2000 {\it Nature} \textbf{407} 355


\bibitem{Jonathan A. Jones} Jones J A, Vedral V, Ekert A and Castagnoli G
2000 {\it Nature} \textbf{403} 869


\bibitem{L.M. Duan} Duan L M, Cirac J I and Zoller P
2001 {\it Science} \textbf{292} 1695

\bibitem{Wang Xiangbin 4} Wang X B and Keiji M
2001 {\it Phys. Rev. Lett.} \textbf{87} 0970901

\bibitem{Zhu} Zhu S L and Wang Z D
2002 {\it Phys. Rev. Lett.} \textbf{89} 097902;
2002 {\it Phys. Rev. A} \textbf{66} 042322

\bibitem{Livermore}  Livermore C,
Crouch C H, Westervelt R M, Campman K L and Gossard A C
1996 {\it Science} \textbf{274} 1332

\bibitem{Zrenner} Zrenner A
Beham E, Stufler S, Findeis F, Bichler M and Abstreiter G
2002 {\it Nature} \textbf{418} 612

\bibitem{Zhu_quant_ph0210027} Zhu S L and Wang Z D
2003 {\it Phys Rev A} \textbf{67} 022319


\bibitem{Norbert Schuch} Schuch N and Siewert J
2002 {\it Preprint} quant-ph/0209035

\bibitem{Makhlin} Makhlin Y, Schon G and Shnirman 1999
{\it Nature} \textbf{398} 305

\bibitem{Lian-ao} Wu L A and Lidar D A
2002 {\it Preprint} quant-ph/0202135

\bibitem{C. Piermarocchi} Piermarocchi C, Chen P, Dale Y S and Sham L J
2002 {\it Phys. Rev.} B \textbf{65} 075307

\bibitem{S. De Rinaldis} Rinaldis S De, 'Amico I D, Biolatti E, Cingolani R and Rossi F
2002 {\it Phys. Rev.} B \textbf{65} 081309

\bibitem{F.J. Rodriguez} Rodriguez F J,
Quiroga L and  Johnson N F 1999 {\it Preprint}
cond-mat/9909139


\end{thebibliography}
\end{document}